\documentclass[conference]{IEEEtran}
\usepackage{dsfont}
\usepackage{cite}
\usepackage{graphicx}
\usepackage{amsmath}
\usepackage{amsfonts}
\usepackage{amssymb}
\usepackage{array}
\usepackage{times}
\usepackage{stfloats}
\usepackage{yhmath}
\usepackage{graphics}
\usepackage{textcomp}
\usepackage{amscd}
\usepackage{epsfig}
\usepackage{psfrag}
\usepackage{rotating}
\usepackage{amsmath}
\usepackage[dvipsnames]{xcolor}
\usepackage{color}
\usepackage{float}
\usepackage{epstopdf}
\usepackage{setspace} 
\usepackage{cuted} 
\usepackage[capitalize]{cleveref}
\usepackage{bm}
\newtheorem{lemma}{Lemma}

\newtheorem{theorem}{Theorem}

\begin{document}
\title{{{Man-in-the-Middle and Denial of Service Attacks in Wireless Secret Key Generation}}}
\author{\IEEEauthorblockN{Miroslav Mitev}

\IEEEauthorblockA{\textit{School of CSEE} \\
\textit{University of Essex}\\
Colchester, UK \\
mm17217@essex.ac.uk}
\and
\IEEEauthorblockN{Arsenia Chorti}
\IEEEauthorblockA{\textit{ETIS, Universit\'e Paris Seine, ENSEA } \\
\textit{Universit\'e Cergy-Pontoise,  CNRS}\\
 Cergy-Pontoise, France \\
arsenia.chorti@ensea.fr}
\and
\IEEEauthorblockN{E. Veronica Belmega}
\IEEEauthorblockA{\textit{ETIS, Universit\'e Paris Seine, ENSEA } \\
\textit{Universit\'e Cergy-Pontoise, CNRS}\\
 Cergy-Pontoise, France \\
belmega@ensea.fr}
\and
\IEEEauthorblockN{Martin Reed}
\IEEEauthorblockA{\textit{School of CSEE} \\
\textit{University of Essex}\\
Colchester, UK \\
mjreed@essex.ac.uk}
}

\maketitle
\begin{abstract}
Wireless secret key generation (W-SKG) from shared randomness (e.g., from the wireless channel fading realizations), is a well established scheme that can be used for session key agreement.  W-SKG approaches can be of particular interest in delay constrained wireless networks and notably in the context of ultra reliable low latency communications (URLLC) in beyond fifth generation (B5G) systems. However W-SKG schemes are known to be malleable over the so called ``advantage distillation'' phase, during which observations of the shared randomness are obtained at the legitimate parties. As an example, an active attacker can act as a man-in-the-middle (MiM) by injecting pilot signals and/or can mount denial of service attacks (DoS) in the form of jamming.  This paper investigates the impact of injection and reactive jamming attacks in W-SKG. First, it is demonstrated that injection attacks can be reduced to -- potentially less harmful -- jamming attacks by pilot randomization; a novel system design with randomized QPSK pilots is presented. Subsequently, the optimal jamming strategy is identified in a block fading additive white Gaussian noise (BF-AWGN) channel in the presence of a reactive jammer, using a game theoretic formulation. It is shown that the impact of a reactive jammer is far more severe than that of a simple proactive jammer.

\end{abstract}

\section{Introduction}
In the past two decades a large number of studies and patents appeared on the topic of wireless secret key generation (W-SKG) schemes that exploit channel reciprocity as the source of shared randomness {(see \cite{Hamamreh18} for a comprehensive review and \cite{Jorswieck15} for a  tutorial on physical layer security including W-SKG)}. Additionally, W-SKG over unauthenticated channels has been proposed in \cite{Maurer03i}. To overcome trivial higher-layer man-in-the-middle (MiM) attacks, as with MiM attacks on unauthenticated Diffie-Hellman schemes, physical layer security technologies have been combined with standard authentication and encryption (AE) schemes \cite{Saiki15}. Furthermore, a large number of practical demonstrators have  provided ``proof of concept'' \cite{Jana13}, \cite{Pierrot13}. A resurgence of interest in W-SKG has been witnessed recently as these technologies could be considered for application in B5G systems \cite{Hamamreh18}, in particular in the context of Internet of things (IoT) \cite{LoRa-key19} and -- potentially -- URLLC applications. W-SKG could be a good fit in these systems as the limited computational resources and strict delay constraints can render challenging the use of standard security protocols such as the transport layer security protocol (TLS) protocol and it's IoT friendly version, the datagram transport layer security (DTLS) protocol.

In recent works it has been shown that building semantically secure AE protocols using the W-SKG procedure is straightforward, as long as the channel probing phase of the scheme is robust against active attacks \cite{Saiki15}, \cite{Chorti18}. Therefore, an important next step is to study MiM and denial of service (DoS) attacks during the channel excitation phase of the W-SKG protocol, commonly referred to as ``advantage distillation'' \cite{Saiki15}.
In this paper, two such active attacks, during channel probing are discussed.

Firstly, MiM attacks, referred to as ``injection'' attacks, are investigated in Section \ref{sec:Injection}: an active adversary tries to control part of the generated secret key by spoofing the channel estimation phase of the W-SKG scheme. Existing works have considered jamming attacks and formulate these in game-theoretic form~\cite{anti-jam1}, \cite{anti-jam2}. However, they have not considered the close relationship between injection and jammming. Here we propose a simple approach to mount such a MiM attack, assuming that the adversary has one additional antenna with respect to the legitimate users. This is a very mild assumption with respect to the adversary's capabilities and reveals a critical  vulnerability of W-SKG, that needs to be addressed. As a countermeasure, we propose a concrete pilot randomization scheme using quadrature amplitude phase shift keying (QPSK) modulated random pilots. We prove that the source of shared randomness remains Gaussian and that the adversary can no longer mount the MiM attack. An interesting conclusion of our analysis is that the MiM injection attack is reduced to a jamming attack when pilot randomization is employed.

Motivated by this result, in Sections \ref{sec:Jamming} and IV,  DoS in the form of reactive jamming is studied for BF-AWGN channels -- used as an abstraction for orthogonal frequency division multiplexing (OFDM) modulation systems. The attacker's optimal strategies are derived. In the present contribution we  assume that the legitimate users blindly adopt a uniform power allocation policy, the level of which we optimally identify; the more general case of an arbitrary power allocation for the legitimate parties will be investigated in the future. 
Our study demonstrates that a reactive jammer can have a far more serious impact on the W-SKG process compared to a simple proactive jammer. In the future, frequency hopping as well as energy harvesting approaches to mitigate the impact of reactive jammers \cite{Veronica16} will be explored. 

Finally, conclusions and further work are discussed in Section V.

\section{MiM in W-SKG Systems: Injection Attacks}\label{sec:Injection}

MiM in the form of injection attacks constitutes one of the most critical limitations in W-SKG systems that extract secret keys from received signal strength (RSS) measurements \cite{jana2009effectiveness,Eberz12,Rong15}. 
Recently, various possible approaches for injection attacks have been published: in \cite{jana2009effectiveness}, the attacker controlled the movement of  objects in an indoor wireless network, thus generating predictable changes in the RSS, (e.g., by obstructing, or not, a line-of-sight). In \cite{Eberz12}, whenever similar channel envelope measurements in the links to the legitimate nodes were observed, the MiM spoofed the W-SKG process by injecting a strong signal.  
In the following we will prove that -- even when full CSI is used to extract the keys -- it suffices that the adversary has one additional antenna with respect to the legitimate users to be able to mount an injection MiM attack. 
\label{sec:system}
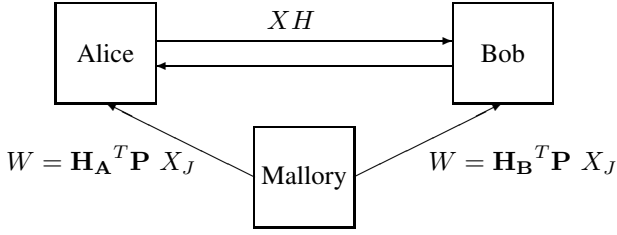
\begin{figure}[t]
\setlength{\unitlength}{0.13in} 
\centering 
\begin{picture}(23.5,11) 
\put(2,6){\framebox(4,4){Alice}}
\put(10,1){\framebox(4,4){Mallory}}
\put(18,6){\framebox(4,4){Bob}}
\put(6,8.5){\vector(1,0){12}}
\put(10.5,9) {{$XH$}}
 \put(18,7.5){\vector(-1,0){12}}
 \put(11,7.5) {}
\put(10,3){\vector(-2,1){6}}
\put(0,3.3) {{$W=\mathbf{H_A}^T \mathbf{P} \ X_J$}}   
\put(14,3){\vector(2,1){6}}
\put(17,3.3) {$W=\mathbf{H_B}^T\mathbf{P} \ X_J$}
\end{picture}
\caption{Alice and Bob have single transmit and receive antennas and exchange pilot signals $X$ over a Rayleigh fading channel with realization $H$. A MiM, Mallory, with multiple transmit antennas can inject a suitably pre-coded signal $\mathbf{P} X_J$, such that the received signal at both Alice and Bob coincide $W=\mathbf{H_A}^T \mathbf{P}= \mathbf{H_B}^T\mathbf{P}$.}
\label{fig:systemmodel}
\end{figure}

To capture the main components of MiM attacks in W-SKG systems, we employ the system model depicted in Fig. \ref{fig:systemmodel}, comprising three nodes: a legitimate transmitter, its intended receiver, and a MiM, referred to as Alice, Bob and Mallory, respectively. Alice and Bob are assumed to have a single antenna each for simplicity, while Mallory has two transmit antennas.\footnote{It is straightforward to see that the scenario can easily be generalized to a multi-antenna setting in which Mallory has one more antenna than Alice and Bob.}
The fading channel realization in the link Alice-Bob is denoted by the complex circularly symmetric Gaussian random variable $H\sim \mathcal{CN}(0, \sigma^2)$. To obtain estimates of $H$, Alice and Bob exchange pilot signals $X$ with $\mathds{E}[|X|^2]\leq P$. Furthermore, following \cite{Eberz12}, we assume Mallory has perfect knowledge of the channel vectors in the multiple input single output (MISO) links Mallory-Alice and Mallory-Bob. The channel coefficients are assumed to be independent and identically distributed (i.i.d.), i.e., $\mathbf{H_A}=[H_{A1}, H_{A2}]^T, \mathbf{H_B}=[H_{B1}, H_{B2}]^T$ with $(H_{A1}, H_{A2}, H_{B1}, H_{B2}) \sim \mathcal{CN}(\mathbf{0}, \sigma_J^2/2\ \mathbf{I}_4)$; this assumption is realistic since Mallory can estimate the channel vectors while Alice and Bob exchange pilot signals, as long as the channel's coherence time is respected (a plausible scenario in slow fading, low mobility environments).  

To mount the attack, Mallory transmits a signal $X_J$, suitably precoded as $\mathbf{P}X_J$. The precoding matrix $\mathbf{P}=[P_1, P_2]^T$ is chosen such that the same signal is ``injected'' at both Alice and Bob, i.e.,
\begin{align}
    \mathbf{H_A}^T\mathbf{P} X_J&=\mathbf{H_B}^T\mathbf{P}X_J
    \Rightarrow\nonumber\\ P_1&=\frac{H_{B2}-H_{A2}}{H_{A1}-H_{B1}}P_2,
\end{align}
where, due to the i.i.d. assumption and to the continuous distribution of the channels, $H_{A1}\neq H_{B1}$ almost surely.
As a result, Mallory can select a suitable precoding matrix (among infinite possibilities). Assuming a total power constraint $\mathds{E}[|\mathbf{P}X_J|^2]\leq \Gamma$ for Mallory's transmission, $P_2$ should be chosen as
\begin{equation}
    P_2\leq \frac{\sqrt{\Gamma}}{\left|\frac{H_{B2}-H_{A2}}{H_{A1}-H_{B1}}\right|}.
\end{equation}

This procedure, illustrated in Fig. 1, shows that it is possible to generalize the injection attack presented in  \cite{Eberz12}, in which an attacker injected a strong signal whenever the RSS in the Mallory-Alice and Mallory-Bob links were similar. More importantly, the presented injection attack accounts not only for the RSS but for the full CSI, i.e., it includes the signal phase.

The observations at Alice and Bob, denoted by $Z_A$ and $Z_B$, are
\begin{align}
Z_A=XH+W+N_A \label{eq:Za}\\
Z_B=XH+W+N_B, \label{eq:Zb}
\end{align}
where $W=\mathbf{H_A}^T\mathbf{P} X_J=\mathbf{H_B}^T\mathbf{P}X_J$ denotes the observed injected signal at Alice and Bob which is identical at both due the precoding matrix $\mathbf{P}$; and, $N_A, N_B$ denote zero-mean unit variance i.i.d. complex circularly symmetric Gaussian random noise variables, i.e., $ N_A, N_B \sim \mathcal{ CN}\left( 0, 1 \right)$. 
The secret key rate controlled by Mallory is upper bounded by \cite{Chorti18}
\begin{equation}
L \leq I(Z_A, Z_B; W). \end{equation}
 Identifying the optimal injection signal $W$, corresponds to finding the capacity achieving input signal of the \textit{two-look Gaussian channel} in (\ref{eq:Za})-(\ref{eq:Zb}). This signal is known to be Gaussian \cite{CoverBook}; hence, a good choice for $X_J$ is to be constant, so that, the overall injected signal is an optimal complex zero-mean circularly symmetric Gaussian signal, $W\sim \mathcal{CN}(0, \sigma_J^2\Gamma)$.

A countermeasure to injection attacks can be built by randomizing the pilot sequence exchanged between Alice and Bob  \cite{Chorti18}, \cite{Rong15}. Here, we propose to randomize the pilots by drawing them from a (scaled) QPSK modulation, as follows: 
instead of transmitting the same probing signal $X$, Alice and Bob transmit independent, random probe signals $X$ and $ Y$, respectively, drawn from i.i.d. zero-mean discrete uniform distributions ${\mathcal{U}(\{\pm r \pm jr\})}, \text{ where } j=\sqrt{-1}, r=\sqrt{P/2}$, so that, $\mathds{E}\left[X\right]=\mathds{E}\left[Y\right]=0$,  $\mathds{E}\left[|X|^2\right]=\mathds{E}\left[|Y|^2\right]=P$ and $\mathds{E}\left[XY\right]=0$, i.e., the pilots are randomly chosen  QPSK signals. Alice's  observation $Z_A$ is modified accordingly as 
\begin{eqnarray}
{Z}_A&=&YH+W+N_A,
\end{eqnarray}
while Bob's observation is given in (\ref{eq:Zb}).
To establish shared randomness in spite of the pilot randomization, Alice and Bob post-multiply $Z_A$ and $Z_B$ by their randomized pilots, obtaining local observations $\tilde{Z}_A$ and $\tilde{Z}_B$ (unobservable by Mallory), expressed as:
\begin{eqnarray}
\tilde{Z}_A&=&X Z_A=XYH+XW+XN_A,\\
\tilde{Z}_B&=&YZ_B=XYH+YW+YN_B.
\end{eqnarray}

\begin{lemma} The source of shared randomness, when the pilots are randomized QPSK symbols, is a circularly symmetric zero mean Gaussian random variable, $XYH \sim C\mathcal{N}(0, P^2\sigma^2)$.
\end{lemma}
\begin{IEEEproof}
We treat the two orthogonal axes (real and imaginary) independently. Looking only at the real values of the pilots and of the channel coefficient $X,Y, H$ denoted here by $X_R=\mathrm{Re}(X)$, $Y_R=\mathrm{Re}(Y)$ and $H_R=\mathrm{Re}(H)$, we express the underlying discrete uniform pdf $f_{X_R}(x)$ and $f_{Y_R}(y)$ and the continuous pdf $f_{H_R}(h)$ as 
\begin{eqnarray}
    f_{X_R}(x)&=&\frac{1}{2}\delta(x-r)+\frac{1}{2}\delta(x+r),\\
    f_{Y_R}(y)&=&\frac{1}{2}\delta(y-r)+\frac{1}{2}\delta(y+r),\\
    f_{H_R}(h)&=& \frac{1}{\sqrt{\pi}\sigma}e^{-\frac{h^2}{\sigma^2}}.
\end{eqnarray}

The pdf of the product $X_R H_R$ is given as
\begin{eqnarray}
f_{X_R H_R}(z)&=&\int_{-\infty}^{\infty}{f_{X_R}(x)f_{H_R}(z/x)\frac{1}{|x|}\mathrm{d}x}
\nonumber\\ &=&\int_{-\infty}^{\infty}{\frac{1}{2\sqrt{\pi}\sigma |x|}\delta(x-r)e^{-\frac{(z/x)^2}{\sigma^2}}\mathrm{d}x}\nonumber\\
&+&\int_{-\infty}^{\infty}{\frac{1}{2\sqrt{\pi}\sigma |x|}\delta(x+r)e^{\frac{-(z/x)^2}{\sigma^2}}\mathrm{d}x}\nonumber\\
&=&\frac{\sqrt{2}e^{-\frac{2z^2}{P\sigma^2}}}{\sqrt{\pi P}\sigma }
\end{eqnarray}
by substituting $r=\sqrt{P/2}$ at the last derivation, i.e., $X_RH_R\sim \mathcal{N}(0, \frac{P\sigma^2}{4})$. A similar result holds for the products involving also the imaginary parts of $X$ and $H$: $X_I H_I$, $X_I H_R$ and $X_R H_I$, so that $XH\sim \mathcal{CN}(0, P\sigma^2)$. Extending this result, we find that $XHY\sim \mathcal{CN}(0, P^2\sigma^2)$. \end{IEEEproof}

Furthermore, due to the fact that $X$ and $Y$ are independent and have zero mean, the variables $XW$ and $YW$ are uncorrelated, circularly symmetric zero-mean Gaussian random variables, and, therefore independent, while the same holds for $XN_A, YN_B$, i.e., $(XW, YW) \sim \mathcal{CN}(\mathbf{0}, \sigma_J^2P\Gamma\mathbf{I}_2)$ and $(XN_A, YN_B) \sim \mathcal{CN}(\mathbf{0}, P\mathbf{I}_2)$. Alice and Bob extract the common key from the modified source of common randomness $XYH$ as opposed to $XH$. On the other hand, since $XW, YW, X N_A, Y N_B$ are i.i.d. complex circularly symmetric Gaussian random variables, the proposed scheme reduces injection attacks to uncorrelated jamming attacks, i.e., using Lemma 1 we get that \begin{equation}
    L\leq I\left(\tilde{Z}_A, \tilde{Z}_B; W   \right)=0.
\end{equation}

\section{Jamming Attacks on W-SKG}\label{sec:Jamming}
Building on the results of the previous section, we next examine in detail the scenario in which Mallory acts as a reactive jammer. Reactive jamming is a stealthy jamming approach in which the jammer first senses the spectrum and jams only when she detects an ongoing transmission. Due to the effectiveness and difficulty to be detected, reactive jammers are considered as  the most harmful \cite{Fang,Spuhler}. Furthermore, as OFDM is used in many actual systems (and will be used at least in the first deployments of 5G), in our analysis we assume a BF-AWGN channel as in \cite{Veronica16}. 
In this context, we assume that Alice and Bob perform W-SKG over a BF-AWGN channel with $N$ parallel blocks (referred to as subcarriers for clarity). The notation introduced in Section II is extended with the introduction of a carrier index $i\in \{1,\ldots,N\}$, i.e., $X_i, Y_i$ denote the randomized pilots on the $i$-th subcarrier, $H_i$ denotes the channel coefficient in the link Alice-Bob,  $W_i$ the signal injected by Mallory on the $i$-th subcarrier and $N_{A,i}, N_{B,i}$ noise variables.
As a reactive jammer, Mallory senses the spectrum and jams a specific subcarrier only when the power on it exceeds a certain threshold $p_{\text{th}}$. Two scenarios are considered: i) when $p_{\text{th}}$ is fixed (determined in essence by the carrier sensing capability of Mallory's receiver); ii) when $p_{\text{th}}$ is variable (its choice forms part of her strategy).

We can reformulate the expressions of Alice's and Bob's local observations on the $i$-th W-SKG subcarrier as follows:
\begin{eqnarray}
\tilde{Z}_{A,i}&=&X_iY_iH_i + X_iW_{i}+X_iN_{A,i}\\
    \tilde{Z}_{B,i}&=&X_iY_iH_i + Y_iW_{i}+Y_iN_{B,i}
\end{eqnarray}
for $i=1, \ldots, N$ with $H_i  \sim \mathcal{CN}(0, \sigma^2)$, $W_i\sim \mathcal{CN}(0, \sigma_J^2\gamma_i)$, $N_{A,i}\sim \mathcal{CN}(0, 1)$, $N_{B,i}\sim \mathcal{CN}(0, 1)$. In this work, we assume that Alice and Bob use the same power $p$ on all pilots, in agreement with common practice during the advantage distillation phase; the more general scenario of an arbitrary power allocation across the subcarriers will be investigated in the future. 
Based on this assumption we have that $\mathds{E}[|X_i|^2]=\mathds{E}[|Y_i|^2]={p}$ with $p \in [0,P]$. 

On the other hand, we let Mallory choose the power allocation vector to maximize the impact of her attack. The power Mallory uses on the $i$-th subcarrier is denoted by $\gamma_i$, so that $\mathds{E}[|W_i|^2]=\sigma_J^2\gamma_i$. Denoting the average available power for jamming by $\Gamma$ and the power allocation of the jammer by $\underline{\bm{\gamma}}=(\gamma_1, \dots, \gamma_N)$, we assume the following short-term power constraint:
\begin{equation}
\underline{\bm{\gamma}} \in \mathbb{R}_+^N,  \quad \sum_{i=1}^N \gamma_i \leq N\Gamma.  \label{con2}
\end{equation}

Assuming that
$H_i$ is uncorrelated with
$H_{A,i}, H_{B,i}$, $i=1,\ldots,N$ and that the pilot randomization approach proposed in Section II is employed, the W-SKG rate ${R}(p , \gamma_i)=I\left (\tilde{Z}_{A,i}; \tilde{Z}_{B,i} \right)$ on the $i$-th subcarrier, can be expressed as a function of $p$ and $\gamma_i, i=1,\ldots,N$ as \cite{Veronica16}:  
\begin{equation}
{R}(p , \gamma_i)= {\log_2\left(1+ \frac{p \sigma^2}{2(1+\gamma_i\sigma_J^2)+\frac{(1+\gamma_i\sigma_J^2)^2}{p \sigma^2}} \right)}. \label{W-SKGjamm}
\end{equation}
Note that the rate in \eqref{W-SKGjamm} is independent of the instantaneous realizations of the fading coefficients; instead, the variations of the channel gains expressed through the variances $\sigma^2, \sigma_J^2$ determine the rate of the secret keys that can be extracted from the wireless medium.
The overall W-SKG sum-rate can then be simply expressed as follows:
\begin{equation}
{C}_K(p, \underline{\bm{\gamma}})=\sum_{i=1}^N {R(p , \gamma_i)}.
 \end{equation}

\section{Optimal Power Allocation Strategies}
Alice and Bob's common objective is to maximize ${C}_K(p, \underline{\bm{\gamma}})$ with respect to (w.r.t.) $p$, while Mallory wants to  minimize ${C}_K(p, \underline{\bm{\gamma}})$ w.r.t. $\underline{\bm{\gamma}}$. Given the opposed objectives, a non-cooperative zero-sum game can be formulated to study the strategic interaction between the legitimate users and the jammer: $\mathcal{G} = ( \{L,J\}, \{\mathcal{A}_\text{L} , \mathcal{A}_\text{J}(p) \}, {C}_K(p, \underline{\bm{\gamma}}) ) $. The game  $\mathcal{G}$ has three components. Firstly, there are two players: player $L$ representing the legitimate users (Alice and Bob are considered to act as a single player) and player $J$ representing the jammer (Mallory). Secondly, player $L$ has a set of possible actions $\mathcal{A}_{L}=[0,P]$ while player $J$'s set of actions is

\begin{equation}
\mathcal{A}_{J}(p) \! = \! \left\{ \begin{array}{ll}
\{(0, \dots, 0 )\} , &  \text{if } p \leq  p_{\text{th}}, \\
\left \{ \underline{\bm{\gamma}} \in \mathbb{R}_+^N | \sum_{i=1}^N \gamma_i \leq N\Gamma \right \}, & \text{if } p> p_{\text{th}}. \label{51}
\end{array}\right.
\end{equation}
At last, ${C}_K(p, \underline{\bm{\gamma}})$, denotes the payoff function of player $L$. 

Due to the fact that Mallory first observes the transmit power of the legitimate users on the subcarriers and then decides which strategy to choose (a consequence of player $J$ being a reactive jammer), we study a hierarchical game in which player $L$ is the leader and player $J$ is the follower. In this hierarchical game, the solution is the Stackelberg equilibrium (SE) -- rather than Nash -- defined as a strategy profile $(p^{\text{SE}},\underline{\bm{\gamma}}^{\text{SE}})$ where player $L$ chooses his optimal strategy first, by anticipating the strategic reaction of player $J$ (i.e., its best response). This can be rigorously written as:
\begin{align}
&p^{\text{SE}} \triangleq  \underset{ p \in \mathcal{A}_L }{\arg \max}  \sum_{i=1}^N {R(p , \underline{\bm{\gamma}}^*(p))},\text{ and }
\underline{\bm{\gamma}}^{\text{SE}} \triangleq \underline{\bm{\gamma}}^{*}(p^{\text{SE}}),
\end{align}
where $\underline{\bm{\gamma}}^*(p)$ denotes the jammer's best response (BR) function to any strategy $p \in \mathcal{A}_L$ chosen by player $L$, defined as follows:
\begin{equation}
\underline{\bm{\gamma}}^*(p) \triangleq \underset{ \underline{\bm{\gamma}} \in \mathcal{A}_{J}(p) }{\arg \min}  \sum_{i=1}^N {R(p , \underline{\bm{\gamma}})}. \label{BRj1}
\end{equation}
We also denote by $\gamma_i^*(p)$ the $i$-th component of $\underline{\bm{\gamma}}^*(p)$. 

\subsection{Stackelberg equilibrium
with fixed $p_{\text{th}}$}

In the following, we evaluate the SE of the game $\mathcal{G}$ assuming that the threshold $p_{\text{th}}$ is predefined and fixed. The case $P \leq p_{\text{th}}$ is trivial as  $\underline{\bm{\gamma}}^{\text{SE}}=(0, \dots, 0)$, whereas, the legitimate users will optimally use the maximum available power so that $(p^{\text{SE}}=P)$. Indeed, because of the badly chosen threshold or low sensing capabilities of Mallory, the legitimate transmission will never be detected on any of the subcarriers and hence will not be jammed. In the following, we assume that: $P>p_{\text{th}}$.

\begin{lemma} \label{lemma1}
The BR of the jammer for any $p\in \mathcal{A}_L$ chosen by the leader defined in (\ref{BRj1}) is the uniform power allocation, such that:
\begin{equation}
\underline{\bm{\gamma}}^*(p)  \triangleq \left\{ \begin{array}{ll}
(\Gamma, \dots, \Gamma ) , &  \text{if } p> p_{\text{th}}, \\
(0, \dots, 0), & \text{if } p\leq p_{\text{th}}. \label{19}
\end{array}\right.
\end{equation}
\end{lemma}
\begin{IEEEproof}
Note that $R(p, \gamma_i)$ is a monotonically decreasing convex function w.r.t $\gamma_i, \ i=1,\ldots, N$ for any $p>0$. We show that the jamming power should be equally distributed on all of the subcarriers. To prove this, we apply Jensen's inequality using $\delta_i>0, \ \sum_{i=1}^{N} \delta_i =1$, so that 
\begin{math}
R\left(p, \sum_{i=1}^{N}\delta_ix_i\right)\leq \sum_{i=1}^{N}\delta_iR(p, x_i). \label{eq:Jensen}
\end{math} 
Substituting $\delta_i=1/N$,  $x_i=\Gamma/b_i$, we get:
\begin{align}
& R\left(p, \sum_{i=1}^{N} \frac{\Gamma}{N b_i} \right) \leq \sum_{i=1}^{N} \frac{1}{N} R \left(p, \frac{\Gamma}{b_i}\right) \nonumber\Rightarrow \\
&NR\left(p,  \frac{1}{N}\sum_{i=1}^{N} \frac{\Gamma}{b_i} \right) \leq \sum_{i=1}^{N} R \left(p, \frac{\Gamma}{b_i}\right). \label{eq:20}
\end{align}
Applying the power constraint $\sum_{i=1}^{N} \Gamma/b_i \leq N\Gamma$ on the LHS of (\ref{eq:20}), for any $p>p_{\text{th}}$ we have:
\begin{align}
NR\left(p, \Gamma\right)&\!<\!\sum_{i=1}^{N} \! R \!\left(p, \frac{\Gamma}{b_i}\right) \nonumber \Rightarrow\! \!\!\!
& C_K(p, (\Gamma, \dots, \Gamma))\! \leq \! C_K (p, \underline{\bm{\gamma}}),
\label{equalproof}
\end{align}
which shows that in order to minimize $C_K$, Mallory has to distribute her power equally on all subcarriers.
\end{IEEEproof}

In light of this result, the W-SKG sum rate can have two forms:
\begin{equation}
{C}_{K}(p,\underline{\bm{\gamma}}^*(p))= \left\{ \begin{array}{ll}
NR(p,\Gamma) , &  \text{if } p> p_{\text{th}}, \\
NR(p,0), & \text{if } p\leq p_{\text{th}}, \label{CKsimplification}
\end{array}\right.
\end{equation}
which simplifies the players' options.
Next, we address the question of how Alice and Bob should choose their power $p$ optimally by considering the actions available to the players in the game at the key points i.e. at $P$ and $p_{\text{th}}$.
\begin{theorem} \label{th3}
Depending on the available power $P$ for W-SKG, player $L$ will either transmit at $P$ or $p_{\text{th}}$ on all subcarriers. The SE point of the game is unique when $P \neq p_{\text{th}}(\sigma_J^2\Gamma  + 1)$ and is given by
\begin{equation}
(p^{\text{SE}}, \underline{\bm{\gamma}}^{\text{SE}}) \!\!=  \!\!\left\{ \!\!\!\! \begin{array}{ll}
 \{(p_{\text{th}},(0, \dots, 0))\}, & \!\!\!\! \text{if } P<p_{\text{th}}(\sigma_J^2\Gamma \!  + \! 1), \\
\{(P,(\Gamma, \dots, \Gamma))\}, & \!\!\!\! \text{if } P>p_{\text{th}}(\sigma_J^2\Gamma \!  + \! 1).
\end{array}\right.
\end{equation}
When $P=p_{\text{th}}(\sigma_J^2\Gamma  + 1)$, the game $\mathcal{G}$ has two SEs: $(p^{\text{SE}}, \underline{\bm{\gamma}}^{\text{SE}}) \in \{(p_{\text{th}},(0, \dots, 0)) , (P,(\Gamma, \dots, \Gamma)) \}$.
\end{theorem}

\begin{IEEEproof}
Given the BR in (\ref{19}) and the simplification in (\ref{CKsimplification}), player $L$ wants to find the optimal $p \in \mathcal{A}_L$ that maximizes: 
\begin{equation}
R(p,\gamma_i^*(p)) =  \left\{ \begin{array}{ll}
R(p,0), &  \text{if } p \leq p_{\text{th}}, \\
R(p,\Gamma), & \text{if } p > p_{\text{th}}. 
\end{array}\right.
\end{equation}
Given that $R(p,\gamma)$ is monotonically increasing with $p$ for fixed $\gamma$, two cases are distinguished: a) $p\in [0,p_{\text{th}}]$, b) $p\in (p_{\text{th}}, P]$. The optimal $p$ in each case is given by

 a) 
\begin{math}
\underset{ p \in [0, p_{\text{th}}] }{\arg \max}  \ {R(p , \gamma_i^*(p))}=\underset{ p \in [0, p_{\text{th}}] }{\arg \max}  \ {R(p , 0)} =p_{\text{th}},
\end{math}

 b) 
\begin{math}
\underset{ p \in (p_{\text{th}}, P ] }{\arg \max} \ {R(p , \gamma_i^*(p))}=\underset{ p \in (p_{\text{th}}, P ] }{\arg \max} \  {R(p , \Gamma)}=P.
\end{math}

\noindent From a) and b), we conclude that the overall solution is $p^{\text{SE}}=$
\begin{equation}
\underset{ p \in \mathcal{A}_L }{\arg \max} \  R(p,\gamma_i^*(p)) =  \left\{ \begin{array}{ll}
\!\!p_{\text{th}}, \!\!\!&  \text{if } R(P,\Gamma) < R(p_{\text{th}},0), \\
\!\!P, \!\!\!& \text{if }  R(P,\Gamma) > R(p_{\text{th}},0), \\
\!\!\{p_{\text{th}},P\},  \!\!\!& \text{if } R(P,\Gamma) = R(p_{\text{th}},0). \nonumber
\end{array}\right.
\end{equation}

To simplify the three possibilities, we focus on the case when transmitting at full power $R(P,\Gamma)$ (hence being sensed and jammed) is equal to the case when player $L$ is transmitting at threshold $p_{\text{th}}$ (the jammer is silent) i.e., $R(P,\Gamma) = R(p_{\text{th}},0)$. Using this equality, and by substituting appropriately into \eqref{W-SKGjamm}, we obtain a quadratic equation in $P$:
\begin{equation}
P^2\!(2 \sigma^2 p_{\text{th}}\!+\!1)\!-\!P(2{p_{\text{th}}}^2 \sigma^2\!+\!2\sigma_J^2\Gamma{p_{\text{th}}}^2 \sigma^2)-(1\!+\!\sigma_J^2\Gamma)^2 {p_{\text{th}}}^2 \!=\! 0,\nonumber \label{eq:quadratic}
\end{equation}
which has a unique positive root equal to $p_{\text{th}}(\sigma_J^2\Gamma+1)$. Given that, the leading coefficient of \eqref{eq:quadratic}: $(2 \sigma^2 p_{\text{th}}+1)\geq0$ and that $P>0$, we can say that the inequalities $R(P,\Gamma) > R(p_{\text{th}},0)$ and $R(P,\Gamma) < R(p_{\text{th}},0)$ are equivalent to $P>p_{\text{th}}(\sigma_J^2\Gamma+1)$ and $P<p_{\text{th}}(\sigma_J^2\Gamma+1)$, respectively.
\end{IEEEproof}

\begin{figure}[tb]
\centering
\includegraphics[clip, trim=3.5cm 9.4cm 4cm 10cm, width=0.48\textwidth]{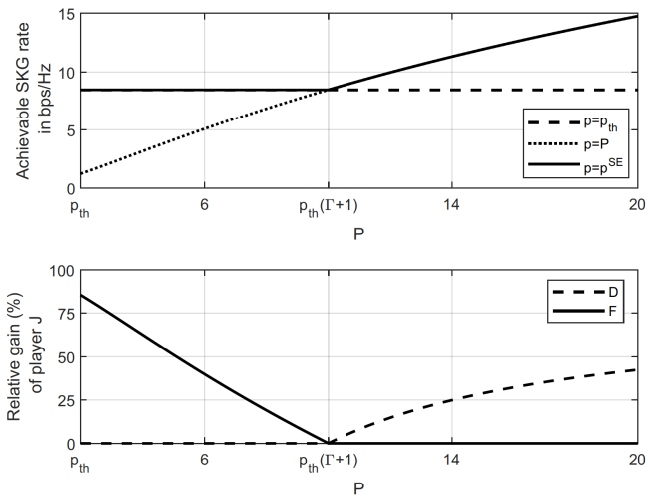}
\caption{UP: SE policy compared to always transmitting with either full power or with $p_{\text{th}}$. DOWN: Functions $D$ and $F$ vs $P$. In both sub-figures, $p_{\text{th}}=2, \Gamma=4, N=10, \sigma^2=\sigma_J^2=1$.}
\label{fig1}
\end{figure}

 \begin{figure}[tb]
 \centering
 \includegraphics[clip, trim=3.3cm 9.4cm 4cm 10cm, width=0.48\textwidth]{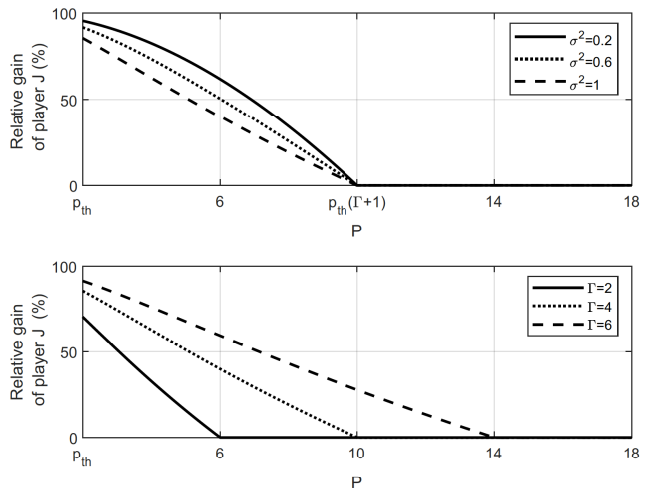}
 \caption{Relative gain of player $J$, evaluated by function $E$, for strategic $p_{\text{th}}$ and fixed $p_{\text{th}}=2$ when $N=10$, $\sigma_J^2=1$ and UP: $\Gamma=4$, DOWN: $\sigma^2=1$.}
 \label{fig2}    
 \end{figure}

Some numerical results are presented in Fig. \ref{fig1} for a total number of SKG subcarriers $N=10$ (pertinent to narrowband IoT applications),  $p_{\text{th}}=2$,  $ \Gamma=4$, and $\sigma^2=\sigma_J^2=1$. The top figure compares the achievable rates of the SE strategy and of two alternative strategies consisting in transmitting with fixed $p=P$ or $p=p_{\text{th}}$. The bottom figure depicts the following quantities:
\begin{align}
F&=\frac{C_K(p^{\text{SE}},\underline{\bm{\gamma}}^{\text{SE}})-C_K(P,(\Gamma, \hdots, \Gamma))}{C_K^{SE}},\\
D&=\frac{C_K(p^{\text{SE}},\underline{\bm{\gamma}}^{\text{SE}})-C_K(p_{\text{th}},(0, \hdots, 0))}{C_K^{SE}},
\end{align}%
where $F$ and $D$ represent the jammer's  gain (or legitimate users' loss) if player $L$ deviates from the SE point (indeed, if player $L$ transmits at $P>p_{\text{th}}$, the jammer will jam at $\gamma_{i}^*(P)=\Gamma$; and if player $L$ transmits at $p_{\text{th}}$ the jammer will not detect it and will remain silent). Both figures show that deviating from the SE point can decrease the achievable sum-rates by up to $85\%$.

\subsection{Stackelberg equilibrium with strategic $p_{\text{th}}$}
Finally, we investigate how Mallory could optimally adjust $p_{\text{th}}$ and how her choice will impact Alice's and Bob's strategies. 
Allowing $p_{\text{th}}$ to vary modifies the game under study as follows $\hat{\mathcal{G}}=(\{L, J\}, \{\mathcal{A}_L , \hat{\mathcal{A}}_J(p)\}, C_K(p,\underline{\bm{\gamma}}, p_{\text{th}}))$, where:
\begin{equation}
\hat{\mathcal{A}}_{J}(p)  \triangleq \left\{ \begin{array}{ll} \! \!\!
\{((0,\dots, 0),p_{\text{th}}), \ p_{\text{th}} \geq 0 \},\!\!\! &  \text{if } p_{\text{th}} \geq  p, \\
\!\!\!\left \{ (\underline{\bm{\gamma}}, p_{\text{th}}) \in \mathbb{R}_+^{N+1} \ | \ \sum_{i=1}^N \gamma_i \leq N\Gamma  \right \},\!\!\! \!& \text{if } p_{\text{th}}< p. 
\end{array}\right.
\end{equation}
The BR of jammer can then be defined as:
\begin{equation}
(\widehat{\underline{\bm{\gamma}}}^*(p), \widehat{p_{\text{th}}}^*(p)) \triangleq  \underset{ (\underline{\bm{\gamma}}, p_{\text{th}}) \in \hat{\mathcal{A}}_J(p) }{\arg \min}  C_K(p,\underline{\bm{\gamma}},p_{\text{th}}).
\end{equation}
\begin{lemma}
The BR of player $J$ in this case is a set of strategies:
\begin{align}
(\widehat{\underline{\bm{\gamma}}}^*(p), {\widehat{p_{\text{th}}}}^*(p))\in \{\  ((\Gamma, \dots, \Gamma), \epsilon), \  \epsilon \in [0,p)\}. \label{BRj2}
\end{align}
\end{lemma}
\begin{IEEEproof}
The problem that the jammer wants to solve is:
\begin{math}
\underset{ (\underline{\bm{\gamma}}, p_{\text{th}}) \in \hat{\mathcal{A}}_J(p) }{\text{min }}  C_K (p,\underline{\bm{\gamma}},p_{\text{th}}),
\end{math}
which can be split as follows:
\begin{align}
\underset{ p_{\text{th}}  \geq 0}{{\text{min}}} \ \ \underset{ \underline{\bm{\gamma}} \in \hat{\mathcal{A}}_J(p) }{{\text{min }}}  C_K (p,\underline{\bm{\gamma}}(p),p_{\text{th}}). \label{minimization}
\end{align}
The solution of the inner minimization is already known from (\ref{19}). For the outer problem we have to find the optimal $p_{\text{th}} \geq 0$ that minimizes $C_K(p,\widehat{\underline{\bm{\gamma}}}^*(p),p_{\text{th}})$. Given that:
\begin{equation}
 \underset{ p_{\text{th}}  \geq 0}{\text{min}}  C_K (p,\widehat{\underline{\bm{\gamma}}}^*(p),p_{\text{th}})\!\!=\!\! \left\{ \begin{array}{ll}
NR(p,\Gamma,p_{\text{th}})  , &  \text{if } p_{\text{th}}<p, \\
NR(p,0,p_{\text{th}}), & \text{if } p_{\text{th}}\geq p,  \label{thres}
\end{array}\right.
\end{equation}
and that $R(p,\Gamma,p_{\text{th}})<R(p,0,p_{\text{th}})$ the jammer can optimally choose any threshold such that $p_{\text{th}}=\epsilon, \ \ \forall \epsilon < p$. meaning, any ongoing transmission is sensed and jammed.
\end{IEEEproof}
  \begin{figure}[tb]
 \centering
 \includegraphics[clip, trim=3.3cm 5cm 4cm 10cm, width=0.48\textwidth]{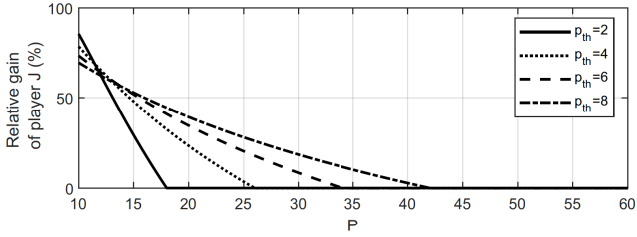}
 \caption{Relative gain of player $J$, evaluated by function $E$, for different values of $p_{\text{th}}$ for $N=10$, $\sigma_J^2=1$ and $\Gamma=4$.}
 \label{fig3}    
 \end{figure}
\begin{theorem}
The game $\hat{\mathcal{G}}$ has an infinite number of SEs:
\begin{equation}
(\widehat{p}^{\text{SE}}, \widehat{\underline{\bm{\gamma}}}^{\text{SE}}, \widehat{p_{\text{th}}}^{\text{SE}})\in \{ \ (P,(\Gamma, \dots, \Gamma),\epsilon), \ \  \forall \epsilon < P\}.
\end{equation}
\end{theorem}
\begin{IEEEproof}
Given the BR of player $J$, we will now evaluate the SE of the game $\hat{\mathcal{G}}$. The definition for $\widehat{p}^{\text{SE}}$ is given as:
\begin{equation}
\widehat{p}^{\text{SE}} \triangleq \underset{p \in \mathcal{A}_L}{\arg } \max C_K (p,\widehat{\underline{\bm{\gamma}}}^*(p),\widehat{p_{\text{th}}}(p)^*).
\end{equation}
Since the jammer will act as in (\ref{BRj2}), we have:
\begin{equation}
 C_K (p,\widehat{\underline{\bm{\gamma}}}^*(p),\widehat{p_{\text{th}}}(p)^*)=NR(p, \Gamma, \epsilon), \ \forall \epsilon < p,
\end{equation}
and the fact that $R(p, \Gamma, \epsilon)$ is monotonically increasing with $p$ results in $\widehat{p}^{\text{SE}}=P$.
\end{IEEEproof}

Fig. \ref{fig2} and Fig. \ref{fig3} illustrate the gain of the jammer (or the loss in W-SKG rate) when $p_{\text{th}}$ is part of her strategy, with utility function $C_K(p,\underline{\bm{\gamma}},p_{\text{th}})$, compared to the case when it is not, with utility function $C_K(p,\underline{\bm{\gamma}})$. We evaluate this gain by:
\begin{equation}
 E=\frac{C_K(p^{SE},\underline{\bm{\gamma}}^{SE})-C_K(\widehat{p}^{SE},\widehat{\underline{\bm{\gamma}}}^{SE},\widehat{p_{\text{th}}}^{SE})}{C_K(p^{SE},\underline{\bm{\gamma}}^{SE})}. \label{eq:E}
 \end{equation}
 As in Fig. \ref{fig1} the total number of subcarriers is $N=10$ and $\sigma_J^2=1$.
 The non-strategic threshold on Fig. \ref{fig2} is set to $p_{\text{th}}=2$ and the quantity $E$ is evaluated for different values of $\sigma^2$ and $\Gamma$. The numerical results demonstrate that when $p_{\text{th}}$ is part of Mallory's strategy, she can be a significantly more effective opponent, compared to the case when $p_{\text{th}}$ is fixed, confirming that reactive jammers can indeed pose a serious threat. This is also confirmed by the results on Fig. \ref{fig3} where the relative gain of the jammer is presented for different $p_{\text{th}}$. As expected with decreasing the threshold her gain increases.

\section{Conclusions}\label{sec:conclusions}
In this study, injection and reactive jamming attacks were analyzed in W-SKG systems and optimal power allocation policies were investigated in BF-AWGN channels. It was shown that pilot randomization can reduce injection MiM attacks to less harmful jamming attacks. An intelligent reactive jammer should optimally jam with equal power on the whole spectrum. Furthermore, a strategically chosen jamming threshold just below the power level used by the legitimate users, allows the adversary to launch a much more effective attack. In this case, the legitimate users have no choice but to transmit at full power.
\section*{Acknowledgements}
M. Mitev is supported by the Doctoral Training Programme of CSEE, University of Essex, A. Chorti and E.V. Belmega are supported by the ELIOT ANR-18-CE40-0030 and FAPESP 2018/12579-7 project and M.J. Reed is supported by the project SerIoT which has received funding from the European Union's Horizon 2020 Research and Innovation programme under grant agreement No 780139.
\bibliographystyle{IEEEtran}
\bibliography{refs}

\end{document}